\documentclass[aip,jcp,%
amsmath,amssymb,%
reprint,
superscriptaddress,%
footinbib,%
]{revtex4-1}

\usepackage{graphicx}
\usepackage{dcolumn}
\usepackage{bm}
\usepackage{hyperref}

\usepackage[usenames,dvipsnames]{color}
\usepackage[normalem]{ulem}

\newcommand{\kbt}[0]{ k_\mathrm{B}T}
\newcommand{\bs}[1]{ \boldsymbol{#1}}

\newcommand{\figref}[1]{Fig.~#1}

\begin{document}

\title[Nudged Elastic Band calculation of the binding potential for liquids at interfaces]{Nudged Elastic Band
  calculation of the binding potential for liquids at interfaces}

\author{Oleg Buller}%
\thanks{O.B. and W.T.  contributed equally to this work.}
\affiliation{%
 Institute for Physical Chemistry, University of M\"unster, Correnstr.~28/30, 48149 M\"unster, Germany
}%

\author{Walter Tewes}
\thanks{O.B. and W.T.  contributed equally to this work.}

\affiliation{%
  Institute for Theoretical Physics, University of M\"unster,  Wilhelm-Klemm-Str.~9, 48149 M\"unster, Germany
}%

\author{Andrew J. Archer}
\email{a.j.archer@lboro.ac.uk}
\affiliation{Department of Mathematical Sciences, Loughborough University, Loughborough LE11 3TU, United Kingdom}

\author{Andreas Heuer}
\affiliation{%
 Institute for Physical Chemistry, University of M\"unster, Correnstr.~28/30, 48149 M\"unster, Germany
}%
\affiliation{
  Center for Nonlinear Science (CeNoS), University of M\"unster, Corrensstr.~2, 48149 M\"unster, Germany
}%
\affiliation{
  Center for Multiscale Theory and Computation (CMTC), University of M\"unster, Corrensstr.~40, 48149 M\"unster, Germany
}
\author{Uwe Thiele}
\email{u.thiele@uni-muenster.de}
\affiliation{%
  Institute for Theoretical Physics, University of M\"unster,  Wilhelm-Klemm-Str.~9, 48149 M\"unster, Germany
}%
\affiliation{
  Center for Nonlinear Science (CeNoS), University of M\"unster, Corrensstr.~2, 48149 M\"unster, Germany
}%
\affiliation{
  Center for Multiscale Theory and Computation (CMTC), University of M\"unster, Corrensstr.~40, 48149 M\"unster, Germany
}%

\author{Svetlana V. Gurevich}
\affiliation{%
  Institute for Theoretical Physics, University of M\"unster, Wilhelm-Klemm-Str.~9, 48149 M\"unster, Germany
}%
\affiliation{
  Center for Nonlinear Science (CeNoS), University of M\"unster, Corrensstr.~2, 48149 M\"unster, Germany
}%
\affiliation{
  Center for Multiscale Theory and Computation (CMTC), University of M\"unster, Corrensstr.~40, 48149 M\"unster, Germany
}%

\begin{abstract}

The wetting behavior of a liquid on solid substrates is governed by the nature of the effective
interaction between the liquid-gas and the solid-liquid interfaces, which is described by
the binding or wetting potential $g(h)$ which is an excess free energy per unit area that
depends on the liquid film height $h$. Given a microscopic theory for the liquid, to
determine $g(h)$ one must calculate the free energy for liquid films of any given value
of $h$; i.e.\ one needs to create and analyse out-of-equilibrium states, since at equilibrium
there is a unique value of $h$, specified by the temperature and chemical potential of the surrounding
gas. Here we introduce a Nudged Elastic Band (NEB) approach to calculate $g(h)$ and illustrate the method by applying
it in conjunction with a microscopic lattice density functional theory for the liquid. We show too
that the NEB results are identical to those obtained with an established method based on using a fictitious
additional potential to stabilize the non-equilibrium states. The advantages of the NEB approach are discussed.

\end{abstract}


\maketitle

\section{Introduction: Relevance of the Binding Potential}

To describe a thin film of liquid on a surface with liquid-gas interface close to
a solid-liquid interface, the so-called binding
potential \cite{hughes2016influence} (also referred to as the wetting or
disjoining potential \cite{pismen2001nonlocal,de1985wetting} or
effective interface potential \cite{schick1990liquids}) is of
great importance. The binding potential $g(h)$ is an excess free energy
per substrate area due to the interaction of the two interfaces. It
depends on the film height $h$, i.e.\ on the distance between the two interfaces.
$g(h)$ is a key quantity in the study of wetting transitions \cite{dietrich88,schick1990liquids}
and is a crucial input to coarse-grained (mesoscopic) effective interface models
which are used to study both the statics and dynamics of liquids at interfaces.
The binding potential is defined for a uniform thickness layer of the liquid
on a flat solid wall in the presence of a bulk vapor phase. For
partially wetting liquids that form droplets on a solid substrate, the binding potential is particularly
important for describing the droplets
in the vicinity of the three-phase contact line.

On the one hand, expressions for $g(h)$ may be derived from microscopic
theories by asymptotic methods \cite{pismen2001nonlocal} resulting
in relatively simple approximations which consist of
combinations of power laws and/or exponentials. \cite{de1985wetting,teds1988rpap}
Such expressions are used in many applications
although strictly speaking they are only valid in the large $h$ limit. In
particular, the divergence of power law terms for vanishing $h$ are
problematic.  On the other hand, one can avoid such problems by
numerically determining $g(h)$ from microscopic models to obtain a
relation that is valid for all film heights. Then, the binding
potential is a useful tool to bridge the scales from a quantitative
microscopic description to a corresponding mesoscopic coarse grained
description where it enters the effective interface Hamiltonian or
mesoscopic free energy. In particular, binding potentials have been
extracted from molecular dynamics (MD) computer simulations
\cite{macdowell2011,tretyakov2013parameter}, from lattice density functional theory
(DFT) \cite{hughes2015142} and continuum DFT
\cite{hughes2016influence}.

\begin{figure}[t]
  \centering
  \includegraphics[width=0.48\textwidth]{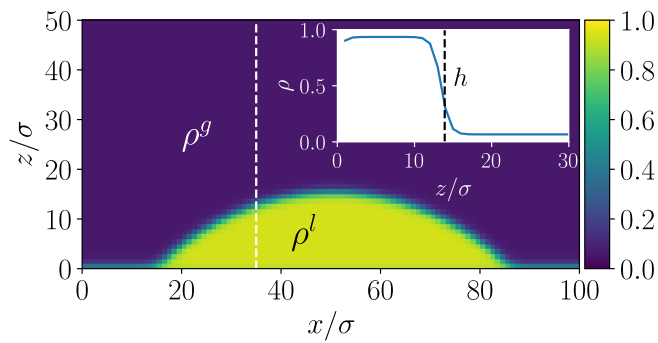}
  \caption{\label{fig:dft_droplet} A droplet density profile obtained from lattice DFT in a
    domain of size $L_x \times L_z = 100 \sigma \times 50 \sigma$
   with lattice spacing $\sigma$ and a wall for $z \le 0$.
   The colour-coding for the density scale is
   given on the right. The inset shows a density profile of a cut through the droplet
   at the position indicated by the white dashed line. The height $h$ is marked with
   a black dashed line at the interface between the liquid ($\rho^{l}$) and gas ($\rho^{g}$) phase.
}
\end{figure}

To illustrate the reasoning leading to the definition of the
binding potential as the contribution to an effective interface
Hamiltonian for a partially wetting liquid on a substrate, we consider
a two-dimensional (2D) system
(cf.~\figref{\ref{fig:dft_droplet}}), with fluid contained in a rectangular domain, $A=[0,L_x]\times[0,L_z]$.
A thermodynamic description of such a three phase system can be done using
DFT \cite{evans1979nature, hansen2013theory} in the canonical ensemble,
i.e., based on the minimization of a Helmholtz free energy
$\mathcal{F}\left[\rho\right]$ as a functional of the density profile,
$\rho(\mathbf{x}),~\mathbf{x}\in A$, subject to the
constraint that the system contains a fixed number of particles
$N=\int_{A}\rho(\mathbf{x})\mathrm{d}\mathbf{x}$ which is enforced
by a Lagrange multiplier which is also the chemical
potential $\mu$. Thus, the equilibrium state of the system
corresponds to the minimum of the thermodynamic \textit{grand potential}
$\Omega=\mathcal{F}-\mu\int_{A}\rho(\mathbf{x})\mathrm{d}\mathbf{x}$.
When there is no substrate wall in the system, the equilibrium state has a uniform density.
However, below the critical temperature,
two-phase coexistence can occur and one observes phase separation into bulk liquid
and bulk gas phases with densities $\rho^l$ and $\rho^g$,
respectively. At coexistence the chemical potential $\mu=\mu_{\mathrm{coex}}$
and the difference between the value of $\Omega$ when the system
contains just a single phase and when there is gas-liquid coexistence, per unit
area (or length in 2D) of the liquid-gas interfaces, gives the
liquid-gas surface tension $\gamma_{\mathrm{lg}}$. Similarly, when there is
a wall present in the system, the excess free energy per wall-area due to
the wall-liquid interface (at $\mu=\mu_{\mathrm{coex}}$) corresponds to the
solid-liquid interfacial tension $\gamma_{\mathrm{sl}}$. The binding potential is extracted
from the difference of the minimized grand potential for an imposed value of the
adsorption $\Gamma=\int_{0}^{L_{z}}[\rho(z)-\rho^g]\mathrm{d}z$ (identical at all points along
 the wall) and the state with $h\to\infty$
(at $\mu=\mu_{\mathrm{coex}}$).

As a result, the minimization of the grand potential
$\Omega[\rho(\mathbf{x})]$ with respect to $\rho(\mathbf{x})$
is reduced to a minimization of
an effective interface Hamiltonian $F[h(x)]$ with respect to a function
$h(x)$ with $x \in [0,L_{x}]$ that describes the position of the
liquid-gas interface (cf.~\figref{\ref{fig:dft_droplet}}). Of course,
neither functional is known exactly, but a good approximation for the latter is
\begin{equation}
 F\left[h\right]=\int\limits_0^{L_{x}}\left[g\big(h(x)\big)+\gamma_{\mathrm{lg}}
 \xi +\gamma_{\mathrm{sl}} \right]\mathrm{d}x.
\end{equation}
This excess free energy contains contributions from the liquid-gas
and solid-liquid interfaces, i.e., the surface tensions
$\gamma_{\mathrm{lg}}$ and $\gamma_{\mathrm{sl}}$, and the
interaction energy of the two interfaces,
i.e., the binding potential $g(h)$. Here
$\mathrm{d}s=\xi \mathrm{d}x$ is the local interface element, i.e.,
$\xi=\sqrt{1+(\partial_x h)^2}$ represents the interface metric.  For
small interface slopes one can make the small-gradient or long-wave
approximation $\xi\approx1+(\partial_x h)^2/2$ often used in gradient
dynamics models on the interface Hamiltonian (aka thin film or
lubrication models). \cite{ordb1997rmp,thie2010jpcm,yin2016films}

The crucial step in the coarse-graining procedure is to
determine the binding potential from
a set of one-dimensional density profiles $\rho(z)$, where $z$ is the perpendicular
distance from the wall, obtained by minimizing
$\Omega[\rho(z)]$ under the constraint of fixed
adsorptions $\Gamma$ \textit{and} under the condition that
$\mu=\mu_{\mathrm{coex}}$, i.e.\ that $\rho(z)\to\rho^g$ for $z\to\infty$, far from the wall. Since one can define
$h=\Gamma/(\rho^l-\rho^g)$, this is equivalent to imposing the film height
$h$. As discussed in Ref.~\onlinecite{archer2011109}
for the somewhat related case of droplet nucleation, the adsorption constraint
condition can not be implemented through a Lagrange multiplier, since this
would shift the chemical potential away from the coexistence value. In
Refs.~\onlinecite{hughes2015142,hughes2016influence} the authors employ an
iterative Picard algorithm for the minimization of the grand
potential, where the adsorption is imposed through a self-consistent
calculation of an additional fictitious potential
$V_{\mathrm{eff}}(z)$. The a priori unknown
$V_{\mathrm{eff}}(z)$ has to be calculated separately for each
adsorption and can be interpreted as an additional space-dependent
external potential that acts mainly within the liquid, i.e.,
$V_{\mathrm{eff}} \rightarrow 0 $ for $z \rightarrow
\infty$.
A detailed discussion of the fictitious potential method can be found in \onlinecite{archer2011109,hughes2015142}.

Here, we present an alternative approach based on the Nudged Elastic
Band method \cite{henkelman:2000, henkelman2002methods, henkelman:2008}, also well known as a geometry
optimization algorithm used to determine chemical reaction paths. It was successfully employed
for a problem similar to that discussed above, namely to determine the free energy barrier for
the nucleation of a liquid drop in a gas phase. \cite{lutsko2008}
Originally, the NEB method was introduced to determine
saddle points on a potential energy landscape as well as corresponding
steepest descent paths (SDP) connecting saddle points and minima.\cite{schlegel11:geom}
As mentioned above, our aim here is to obtain from DFT the
minimum of the grand potential for a specified adsorption and bulk density.
The SDP on the free energy landscape (with respect to the Euclidean metric) obtained by the NEB method can
be parametrized by the adsorption. The free energy values along the
path are interpreted as an approximation for the required constrained free energy
minima. We compare our results obtained from the NEB method to the corresponding
results obtained via the fictitious potential approach and show that they are in excellent
agreement. Furthermore, we verify that the NEB method does in fact give the SDP by
additionally comparing to results obtained from a pseudo dynamics (i.e. the trajectory
given by a non-conserved dynamical equation based on the DFT).
However, we first introduce the DFT.

\section{Model system}
As a simple model system we consider the
following discrete lattice DFT grand potential in a reduced one-dimensional
(1D) description of a 2D or 3D system
(i.e.\ assuming translation invariance along the wall)
\begin{eqnarray}
  \label{eq:model_system}
  \Omega ( \{ \rho_i \})
  \nonumber            = & \kbt \sum\limits_{i=1}^{L_z} [\rho_i \ln(\rho_i) + (1-\rho_i)\ln(1-\rho_i)] \\
                         &  - \frac 1 2 \sum\limits_{i=1}^{L_z} \sum\limits_{j=1}^{L_z}  \epsilon_{ij} \rho_i \rho_j
                            +\sum\limits_{i=1}^{L_z} \rho_i(V_i - \mu).
\end{eqnarray}
Here, $\{\rho_i\}$ are the densities in a system of size ${L_z}$,
$k_{\mathrm{B}}$ is the Boltzmann constant, $T$ is the temperature (in the following $k_{\mathrm{B}}T=\beta^{-1}$),
$\mu$ is the chemical potential, $\{V_i\}$ is an
external potential modeling the wall interaction, whereas the fluid inter-particle interaction is modeled through the interaction matrix
$\epsilon_{ij}$. In our case, we consider nearest neighbor
(i.e., short-range) interactions as well as particle self-interactions
that results from the mapping of the particle pair interactions in
the full translation-invariant 2D or 3D system, as described in Ref.~\onlinecite{hughes2014introduction}.
In contrast, the wall potential is long-ranged and acts across the entire system, algebraically decaying as
\begin{equation}
V_i= -\epsilon_\omega i^{-3} ~~\mathrm{for}~ i\ge 1,
\end{equation}
i.e., implicitly $V_i=\infty$ for $i<1$.
As a measure of the thickness of the wetting film we define the adsorption on a domain of length ${L_z}$ according to:
\begin{eqnarray}
  \label{eq:adrob_rate}
  \Gamma = \sum_{i=1}^{L_z} \left ( \rho_i - \rho^g  \right )
\end{eqnarray}
with the resulting effective film height
$h=\Gamma/(\rho^l-\rho^g)$.

\section{Nudged Elastic Band approach}

The NEB method belongs to the class of
Double-Ended Chain-of-States methods for geometric optimization
problems \cite{henkelman:2000, henkelman2002methods, henkelman:2008,
  schlegel11:geom}, i.e., on a free energy landscape a set of $P$ points
is distributed between two fixed end points in such a way that a
pre-defined criterion is optimized. Here, these points are represented
by a sequence of density profiles $\bs{\rho}^{I}=\{\rho^{I}_i\},~ I=1 \dots P$, with
corresponding values of the grand potential
$\Omega[ \boldsymbol{\rho}^{I} ]$ and adsorption $\Gamma^{I}$.
As end points ($\boldsymbol{\rho}^0$, $\boldsymbol{\rho}^{P+1}$) of
the chain, {we use the density profiles of a homogeneous
bulk gas ($\boldsymbol{\rho}^0=\boldsymbol{\rho}^g$) at zero adsorption
$\Gamma^0 = 0$ and a bulk liquid phase in contact with the wall
($\boldsymbol{\rho}^{P+1}=\boldsymbol{\rho}^{wl}$) with a corresponding
$\Gamma^{P+1}=\Gamma^{wl}$,} that  minimizes the
unconstrained $\Omega [\boldsymbol{\rho}]$.
As initial guesses for all intermediate profiles
we use hyperbolic tangent functions
\begin{equation}
  \label{eq:initial_set}
  \rho^{I}_i = \frac 1 2 (\rho^g - \rho^l)\mathrm{tanh} \left (i\sigma-a_{I}{L_z} \right ) + \frac 1 2,
\end{equation}
placing the liquid-gas interface at the desired positions via the
parameter $a_I$.

An individual optimization process
for each single $\Omega [ \bs{\rho}^{I} ]$ point on the free energy landscape will
always reach a minimum of the functional. To prevent this and to well distribute the intermediate
profiles $ \bs{\rho}^{I} $ between the two end points, an artificial elastic
force $\bs{F}^{I}_{\text{elastic}}$ is introduced between
profiles. Then
the entire chain of elastically joined profiles is optimized in
nparallel. The elastic energy is
proportional to the squared distance between two neighboring density
profiles (based on the $L_2$ norm)
\begin{equation}
  \label{eq:norm}
| \boldsymbol{\rho}^{I+1} - \boldsymbol{\rho}^{I} |^2 = \sum_{i=1}^{L_z} \left ( \rho^{I+1}_i - \rho^{I}_i \right )^2.
\end{equation}
Note that if we replace the distance measure between profiles in Eq.\ \eqref{eq:norm}
  by $\sum_{i=1}^{L_z} \left | \rho^{I+1}_i - \rho^{I}_i \right |$ or even the difference between
  adsorptions $|\Gamma^{I+1}-\Gamma^I |$, then the results discussed below do not change.
The elastic force only acts along the chain, i.e., it is a parallel
force component, whereas the component perpendicular to the chain, results from the energy functional
$\bs{F}_{\text{FE}}^I = - \partial \Omega( \bs{\rho})/ \partial  \bs{\rho} |_{\bs{\rho} = \bs{\rho}^I} $.
Thus, on the discretized space of density profiles, we employ the Euclidean metric in order to approximate
the constrained minimum free energies by points on a SDP.
The overall force vector $\boldsymbol{F}_{\text{NEB}} = \{\bs{F}_{\text{NEB}}^1,\dots,\bs{F}_{\text{NEB}}^P \}$ to be minimized is composed of
the forces acting at all intermediate points $I=1 \dots P$ (i.e. excluding the end points), where
\begin{equation}
  \label{eq:neb_force}
  \boldsymbol{F}_{\text{NEB}}^I = \underline{\bs{P}}^{\parallel} \boldsymbol{F}_{\text{elastic}}^I + \underline{\bs{P}}{^{\perp}}\boldsymbol{F}_{\text{FE}}^I,
\end{equation}
with $\underline{\bs{P}}^{\parallel}$ and $\underline{\bs{P}}{^{\perp}}$ as projection operators
onto the direction parallel and perpendicular to the chain, respectively. The projection operators
are obtained via a tangent formalism and the detailed form of $\bs{F}^{I}_{\text{elastic}}$
can be found in Ref.~\onlinecite{henkelman:2000}.
For the minimization of $|\boldsymbol{F}_{\text{NEB}}|$ we employ
geometric optimization using direct inversion in the interactive
subspace (GDIIS), see Refs.~\onlinecite{csaszar1984, b108658h}.  Ideally,
this yields a set of profiles $\bs{\rho}^{I}$ that are homogeneously
distributed along the SDP between
$\bs{\rho}^g$, $\bs{\rho}^{wl}$ and the extrema in between. Since the force
vector results from an artificial projection procedure,
it is not obvious that this method is guaranteed to give the exact
SDP. However, in practice it seems to do very well, as is illustrated next.

\section{Comparison of fictitious potential and NEB methods}

\begin{figure*}[t]
  \centering
  \includegraphics[width=1.08\textwidth]{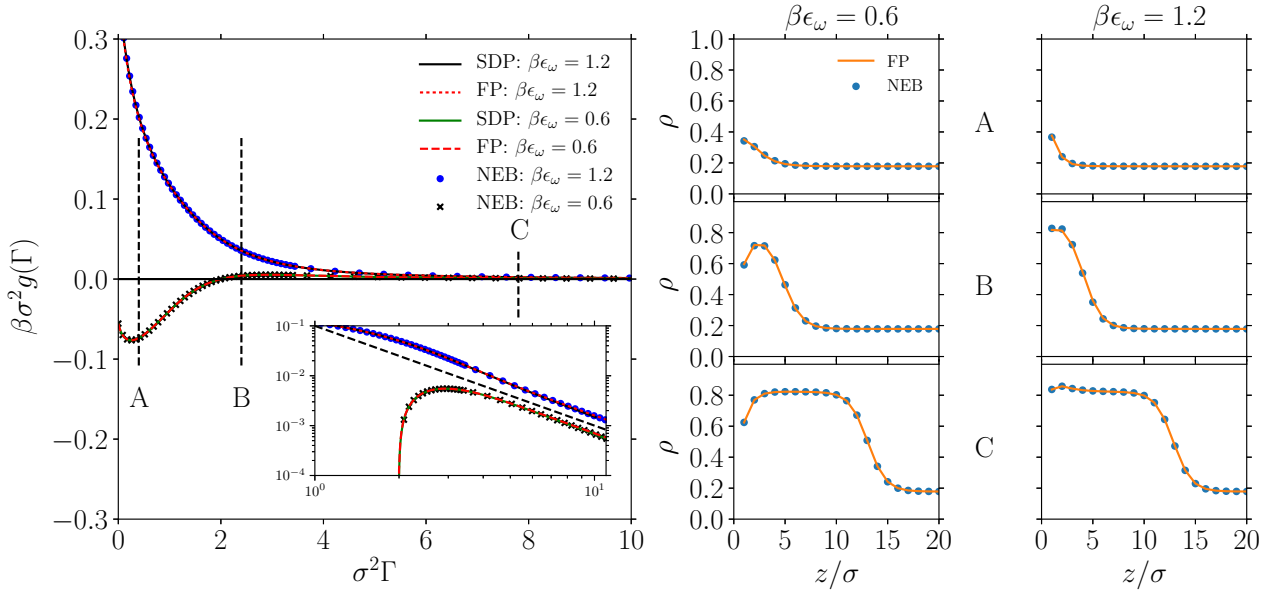}
  \caption{\label{fig:binding_pot_comp} Left: Comparison of the
    binding potentials obtained as a function of the adsorption $\Gamma$, as
    determined by three different methods: fictitious potential method
    (FP), Nudged Elastic Band method (NEB) and  via a pseudo dynamics that
    follows the steepest descent (SDP) in the energy landscape
    initiated at $\bs{\rho}^g$ and the local maximum (for
    $\beta \epsilon_\omega = 0.6$). Two different wall attraction
    strengths are considered, representing the cases of
    wetting ($\beta \epsilon_\omega = 1.2$) and partial wetting ($\beta \epsilon_\omega = 0.6$).
    The inset displays the data in a double logarithmic plot to highlight the
    power law behavior at large film height (the black
    dashed line shows $0.1\Gamma^{-2}$ for comparison). Right: The
    corresponding density profiles are shown for adsorptions
    $\sigma^2 \Gamma \approx 0.4, 2.4$, and $7.8$, which are marked
    in the left panel
    by vertical dashed lines and the letters A, B, and C, respectively.
    }
\end{figure*}

We use the lattice DFT \eqref{eq:model_system} with the
interaction parameters $\beta \epsilon_{ij} = \beta \epsilon = 0.95$,
the chemical potential fixed at coexistence
$\mu=\mu_{\mathrm{coex}} = - 5\epsilon/2$ (cf. Ref.~\onlinecite{hughes2014introduction}) and calculate the binding
potentials for two different values of the wall attraction strength parameter
$\beta \epsilon_\omega$, employing both the fictitious potential
method \cite{hughes2015142} and the NEB
method described in the previous section.
\figref{\ref{fig:binding_pot_comp} (left)} shows that the resulting
binding potentials agree very well, while \figref{\ref{fig:binding_pot_comp} (right)}
illustrates the agreement of the obtained density profiles.

Varying the wall attraction strength parameter $\epsilon_\omega$,
we see agreement for two different wetting regimes, namely,
partial wetting at $\beta \epsilon_\omega = 0.6 $ and complete
wetting at $\beta \epsilon_\omega = 1.2$ -- see Ref.\ \onlinecite{hughes2015142}
for more on the behavior at the wetting transition. We emphasize that at large
adsorption values, both methods
yield the power law decay $g(\Gamma)\sim\Gamma^{-2}$ expected in a system with
long-range interactions \cite{hughes2015142,hughes2016influence,dietrich88, Dietrich1991analytic, schick1990liquids}
(here the interaction with the wall),
cf.~the inset of \figref{\ref{fig:binding_pot_comp}}. Note that in the NEB approach
the chain of states is non-equidistant in $\sigma^2 \Gamma$: the point density is increased in the small adsorption value regime, to obtain a
finer resolution where $g(\Gamma)$ is more strongly varying.

In addition to the two methods introduced above, we calculate a
SDP employing a pseudo-dynamics that starts at $\Omega\left[\boldsymbol{\rho}^g\right]$ {and at both sides of the local maximum for the partial wetting case (the starting points are obtained by the NEB or the fictitious potential method) and}
follows the steepest gradient in small steps to $\Gamma \rightarrow
\infty$ {or to the minimum corresponding to a small finite value of the adsorption at the wall}. The resulting path also agrees perfectly with the NEB
approach, which is what should be expected if both methods approximate the SDP.

\section{Conclusion}
\label{sec:conclusion}

We have presented a Nudged Elastic Band method for calculating the
coarse-grained mesoscopic binding potential $g(h)$ for a liquid film on
a solid substrate based on microscopic DFT. The examples we have considered
indicate that the method yields results that are indistinguishable
from those using the fictitious potential
method of Refs.~\onlinecite{hughes2016influence,hughes2015142}, which is an
approach based on a self-consistent calculation of the fluid density profiles
for specified values of the adsorption. Binding
  potentials obtained with the fictitious potential method were used
  in Ref.~\onlinecite{hughes2016influence} to calculate drop profiles that
  agree remarkably well with drop profiles calculated directly using DFT.
The agreement of fictitious potential and NEB approach represents an important
independent validation of the previous results obtained via the fictitious
potential approach which have already recently been employed in mesoscopic
gradient dynamics models to compute the spreading dynamics of droplets
\textit{and} the advance of adsorption layers by combining advective
and diffusive dynamics.\cite{yin2016films}

Since determining the binding potential involves calculating
non-equilibrium states at $\mu=\mu_{\mathrm{coex}}$ through a constrained
minimization which cannot be formulated in terms of Lagrange
multipliers, it is not a priori obvious that the different approaches
should give identical results. The fact that the different approaches
  are in excellent agreement validates both methods, though we have not rigorously
  proved that our NEB approach gives the desired constrained minimum
  free energies.\footnote{In the context of studying nucleation pathways,
    in Ref.~\onlinecite{Lutsko2012nucleation} a non-Euclidean metric motivated by
    dynamical considerations is employed to calculate a nucleation pathway (in contrast
    to Ref.~\onlinecite{lutsko2008} where a Euclidean metric is considered). The approach
    here is in the first instance applicable to static droplets, so there is no analogous
    dynamical reasoning.}\nocite{Lutsko2012nucleation}

Besides validating the approach of Refs.\ \onlinecite{hughes2016influence,hughes2015142}, our present NEB method represents,
from a technical point of view, a very efficient way to calculate binding potentials that additionally
allows for an intuitive geometric interpretation of the binding
potential as `tracking' the Euclidian steepest descent of the grand potential.

The approach can be readily used for calculations based on other
microscopic models, e.g., continuum DFT models, to derive
coarse-grained mesoscopic binding potentials. Using these in interface
Hamiltonians allows for studies of wetting transitions, mesoscopic
static droplet shapes and even mesoscopic droplet dynamics. It is a
very efficient tool for parameter studies because the entire binding
potential is calculated in one single optimization procedure and the
point chains (chains of density profiles) obtained for one parameter
set can be directly employed as initial guesses for the calculation at
neighboring parameter sets. A future challenge consists in extending
the presented method to determine binding potentials for complex
fluids where it depends not only on film height but also on other,
internal, degrees of freedom. This would allow one to provide
mesoscopic gradient dynamics models for such fluids
\cite{thtl2013prl,thap2016prf} with correct coarse-grained wetting
energies.

\begin{acknowledgments}
  This work was supported by the Deutsche Forschungsgemeinschaft within the
  Transregional Collaborative Research Center TRR 61.
\end{acknowledgments}

\end{document}